\documentclass[12pt]{lmcs}

\keywords{polymorphism, Maltsev operation, relational structure, Datalog programs, constraint satisfaction problem, computational complexity}
\usepackage[numbers,square,comma]{natbib}
\usepackage{float}

\usepackage[nointegrals]{wasysym}
\usepackage{amsmath}
\usepackage{amssymb}
\usepackage{amsthm}
\usepackage{amsaddr}
\usepackage{hyperref}
\usepackage{graphicx}
\usepackage{geometry}
\usepackage{tikz}
\usepackage{tkz-graph}
\usepackage{chngcntr}
\usepackage{thmtools}
\usepackage[linesnumbered,lined,boxed,commentsnumbered]{algorithm2e}
\usepackage{thm-restate}
\usepackage{tabularx}
\usepackage{environ}
\usepackage{algorithm2e}
\usetikzlibrary{shapes}

\counterwithin{figure}{section}

\begin{document}

\title[Existence of Maltsev Polymorphisms]{Complexity of Deciding Existence of Maltsev Polymorphisms of Finite Relational Structures}
\author[D.Delic]{Dejan Delic}
\author[A.Syed]{Ali Syed}
\address{Department of Mathematics, Toronto Metropolitan University}
\email{ddelic@torontomu.ca, ali.mortoza.syed@torontomu.ca}

\newcommand{\A}{\mathbb{A}}
\newcommand{\B}{\mathbb{B}}
\newcommand{\C}{\mathbb{C}}
\newcommand{\D}{\mathbb{D}}
\newcommand{\CSP}{\operatorname{CSP}}
\renewcommand{\t}{\theta}
\newcommand{\Sg}{\operatorname{Sg}}

\begin{abstract}
   \noindent In this article, we show that the problem of deciding whether a finite relational structure $\mathbf{A}$ admits a Maltsev polymorphism is a computational problem, which can be solved in polynomial time. We provide a proof of this result by reducing the problem to deciding whether a binary relational structure, associated with $\mathbf{A}$ admits such a polymorphism. In our construction of such an algorithm, we make an extensive use of a property of binary relations, which is implied by the existence of a Maltsev polymorphism, the so-called rectangularity property. 
\end{abstract}

\maketitle

\section{Introduction}

It was shown in \cite{chen2017asking} that deciding if a relational structure admits any of the following polymorphism is \textsc{NP}-complete; a semilattice polymorphism, a conservative semilattice polymorphism, a commutative semigroup polymorphism. Relational structures and digraphs with majority/ near unanimity polymorphism have studied in \cite{brewster2008near}, \cite{chen2017asking}, \cite{feder2013graphs}, \cite{hell2011dichotomy}, \cite{kazda2011maltsev}, and \cite{maroti2008existence}. However, deciding whether a relational structure $\mathbf{A}$ admit a Maltsev polymorphism seemed to be a more challenging problem. On the algebraic side, the complexity of deciding whether an idempotent algebra has Maltsev term has been investigated (\cite{freese2009complexity}, \cite{horowitz2013computational}, \cite{mayr2012subpower}, \cite{siggers2010strong}). However, a polynomial procedure to recognize whether, given as its input relational structure, that structure has a Maltsev polymorphism remained unknown. Establishing the presence of Maltsev polymorphisms would, for instance show that any instance of the constraint satisfaction problem having  as $\mathbf{A}$ as its template, is solvable in polynomial time (see \cite{bulatov2006simple}).

In this article, we will construct such an algorithm which does not rely on any assumptions on the given finite relational structure $\mathbf{A}$. In particular, we do not work under the so-called ``promise of a Maltsev operation." As will become clear in the paper, the key ingredient in the construction of the said algorithm is the \emph{rectangularity property} of the product of two Maltsev algebras. The failure of rectangularity will immediately preclude the existence of a Maltsev polymorphism of $\mathbf{A}$ and this property can be verified in deterministic logarithmic space in $n=|A|$. 

Once rectangularity of all constraints (binary relations) has been verified, any binary constraint between two simple subuniverses of the domains, which constitutes a subdirect product of the said simple algebras, is either a full product or the graph of an isomorphism between the two simple algebras. After that, we can try to emulate the algorithm from \cite{delic2026maltsev}. The construction from \cite{delic2026maltsev} cannot be fully replicated, since the polymorphisms of $\mathbf{A}$ are not given, and the catalog of possible simple (in fact, minimal) nontrivial subalgebras of the algebra of all polymorphisms of $\mathbf{A}$ is unknown. In \cite{delic2026maltsev}, it is proved that the algorithm presented there, which solves an instance of a Maltsev CSP over the given structure, is in the complexity class $\#$\textsc{L}, the class of problems that are in deterministic logspace with counting. In that algorithm, counting the number of elements of a fixed minimal subalgebra satisfying a particular graph connectivity property played the crucial role. Here, we cannot place the problem in a complexity class, which is conjectured to be strictly below \textsc{P} in polynomial time hierarchy; the algorithm we present in this article is a polynomial time one, which is the sharpest bound we are able to obtain at this point.
    
\section{Background}

\subsection{Maltsev Algebras}

    An \emph{algebra} is an ordered pair $\mathbb{A}=(A, F)$, where $A$ is a nonempty set, the \emph{universe} of $\mathbb{A}$, while $F$ is the set of \emph{basic operations} of $\mathbb{A}$, consisting of functions of arbitrary, but finite, arities on $A$. The list of function symbols and their arities is the \emph{signature} of $\mathbb{A}$. 

    A \emph{subuniverse} of the algebra $\mathbb{A}$ is a nonempty subset $B\subseteq A$ closed under all operations of $\mathbb{A}$. If $B$ is a subuniverse of $\mathbb{A}$, by restricting all operations of $\mathbb{A}$ to $B$, such a subuniverse is a \emph{subalgebra} of $\mathbb{A}$, which we denote $\mathbb{B}\leq \mathbb{A}$. In particular we will be interested in subalgebras generated by a two element subset of $A$. For two distinct elements $a,b \in A$ we use $\Sg(a,b)$ to refer to the smallest subalgebra of $\A$ which contains both $a$ and $b$.

If $\{\mathbb{A}_i\}_{i \in I}$ is an indexed family of algebras of the same signature, the product $\prod_{i \in I} \mathbb{A}_i$ of the family is the algebra whose universe is the Cartesian products of their universes endowed with the basic operations which are coordinatewise products of the corresponding operations in $\mathbb{A}_i$. If $\mathbb{A}$ is an algebra, its $n$-th power will be denoted $\mathbb{A}^n$.

An equivalence relation $\alpha$ on the universe $A$ of an algebra $\mathbb{A}$ is a \emph{congruence} of $\mathbb{A}$, if $\alpha \leq \mathbb{A}^2$, that is to say, if $\alpha$ is preserved by all basic operations of $\mathbb{A}$. In that case, one can define the algebra $\mathbb{A}/\alpha$, the \emph{quotient of} $\mathbb{A}$ \emph{by} $ \alpha$, with the universe consisting of all equivalence classes (cosets) in $A/\alpha$ and whose basic operations are induced by the basic operations of $\mathbb{A}$. The $\alpha$-congruence class containing $a\in A$ will be denoted $a/\alpha$.

An algebra $\mathbb{A}$ is said to be \emph{simple} if its only congruences are the trivial, diagonal relation ($0_\mathbb{A}=\{(a,a)\mid a\in A\}$) and the full relation ($1_\mathbb{A}=\{ (a,b)\mid a,b\in A\}$). It is a well-known fact (see e.g. \cite{Burris1981}) that the congruences of $\mathbb{A}$ form a lattice $Con(\mathbb{A})$; namely, for any $\alpha,\beta\in Con(\mathbb{A})$, $\alpha\wedge\beta$ is the intersection of $\alpha$ and $\beta$, while $\alpha\vee \beta$ is the smallest congruence containing both $\alpha$ and $\beta$.


If $\mathbb{A}$ and $\mathbb{B}$ are two algebras of the same signature, a mapping from $A$ to $B$ which preserves all basic operations is a \emph{homomorphism}. An \emph{isomorphism} is a bijective homomorphism between two algebras of the same signature.

Given an algebra $\mathbb{A}$, a \emph{term} is a syntactical object describing a composition of basic operations of $\mathbb{A}$. A \emph{term operation} $t^\mathbb{A}$ of $\mathbb{A}$ is the interpretation of the syntactical term $t(x_1,\ldots,x_m)$ as an $m$-ary operation on $A$, according to the formation tree of $t$. 
  
A ternary operation $m:A^3\rightarrow A$ on a finite set is said to be \emph{Maltsev} if it satisfies the algebraic identities $m(x,x,y)\approx m(y,x,x)\approx y.$ An algebra $\mathbb{A}=(A, F)$ is said to be a \emph{Maltsev algebra}, if its set of operations $F$ contains a Maltsev operation. 

Clearly, any Maltsev operation $m(x,y,z):A^3\rightarrow A$ is idempotent, since it satisfies $m(x,x,x)\approx x$, for all $x\in A$. Therefore, every singleton $\{x\}$ is a subalgebra of $\mathbb{A}=(A;\{m(x,y,z)\})$. 

Finally, we state some facts about subdirect products of Maltsev algebras which will be needed later. The first fact concerns the connectivity in subdirect products of simple Maltsev algebras as bipartite graphs (i.e. for an algebra $\mathbb{C} \leq_{sp} \mathbb{A} \times \mathbb{B}$ the graph with vertex set $A \cup B$ and edges of the form $xy$ if and only if $xy \in C$). For the proof, see e.g. \cite{Burris1981}

\begin{thm} \label{maltsev} Let $\mathbb{A}_1,\ldots,\mathbb{A}_n$ be simple algebras in a Maltsev variety. If
\[\mathbb{B}\leq_{sp} \mathbb{A}_1\times\ldots\times\mathbb{A}_n\]
is a subdirect product, then
\[\mathbb{B}\cong \mathbb{A}_{i_1}\times\ldots\times\mathbb{A}_{i_k}\]
for some $\{i_1,\ldots,i_k\}\subseteq \{1,\ldots,n\}$. 

In particular, if $\mathbb{A}$ and $\mathbb{B}$ are two simple Maltsev algebras then any subdirect product $\mathbb{C}\leq_{sp} \mathbb{A}\times\mathbb{B}$ is either the direct product or the graph of an isomorphism $f:\mathbb{A}\rightarrow\mathbb{B}$.
\end{thm}

The existence of a Maltsev operation implies the following property on any subdirect product, which we will refer to as the \emph{rectangularity property}:

\begin{prop}\label{rect} Let $C\leq_{sp} \mathbb{A}\times\mathbb{B}$, where $\mathbb{A}$ and $\mathbb{B}$ are algebras in a Maltsev variety. Then, the following holds: if $(a,b), (a,b'), (a',b')\in C$, then $(a',b)\in C$.
\end{prop}

\begin{proof} Let $m(x,y,z)$ be a Maltsev operation on both $\mathbb{A}$ and $\mathbb{B}$. Then,
\[(a',b)=(m(a,a,a'), m(b,b',b'))\in C.\]
\end{proof}




For further facts from universal algebra we direct the reader to~\cite{Burris1981}.

\subsection{Constraint Satisfaction Problem}\label{CSP}

The notion of a constraint satisfaction problem provides a natural framework for a variety of problems which require simultaneous satisfiability of a number of conditions on a given set of variables. More formally, An \emph{instance} of the CSP is a triple $\mathcal{P}=(V,A,\mathcal{C})$, where $V=\{x_1,\ldots,x_n\}$ is a finite set of \emph{variables}, $A$ is a finite domain for the variables in $V$, and $\mathcal{C}$ is a finite set of \emph{constraints} of the form $C=(S,R_S)$, where $S$, the \emph{scope} of the constraint, is a $k$-tuple of variables $(x_{i_1},\ldots,x_{i_k})\in V^k$ and $R_S$ is a $k$-ary relation $R_S\subseteq A^k$, called the \emph{constraint relation} of $C$. A \emph{solution} for the instance $\mathcal{P}$ is any assignment $f:V\rightarrow A$ such that, for every constraint $C=(S,R_S)$ in $\mathcal{C}$, $f(S)\in R_S$.

From this point on, we will  assume that the set of variables $V$ for an instance is an initial segment of the set of positive integers, i.e. $V=\{1,2,\ldots,n\}$, for some $n\geq 1$. A relational structure $\mathbf{A}=(A, \mathcal{R})$, defined over the domain $A$ of the instance $\mathcal{P}$, where $\Gamma$ is a finite set of relations on $A$, is often referred to as a \emph{relational template}, and the relations from $\Gamma$ form the signature of $\mathbf{A}$. An instance of $\CSP (\mathbf{A})$ will be an instance of the CSP such that all constraint relations belong to $\mathbf{A}$.

 \subsection{Existence of Polymorphisms via CSPs}\label{sec:polym}

\noindent In this subsection, we show that, for a fixed $n$, where $n\geq 2$,  all $n$-ary polymorphisms of a fixed finite relational structure $\mathbf{A}$ are precisely the solutions of a particular instance of a constraint satisfaction problem whose domain is $A$.  $n$-ary polymorphisms satisfying particular equational properties (e.g. Maltsev polymorphisms, near unanimity polymorphisms, Taylor polymorphisms, among others) can then be viewed as the solutions of modifications of this canonical problem, obtained by, say, restricting domains of certain variables, introducing binary relations among domains which correspond to the equality relation, etc.

Let $\mathbf{A}=(A; \mathcal{R})$ be a fixed finite relational structure wirth the domain $A$ and a finite set of finitary relations $\mathcal{R}$ on $A$. We can assume that the arities of all relations in $\mathcal{R}$ are bounded above by some fixed $k$, $k\geq 2$.

We next proceed to define the instance $\mathcal{P}_3$ of the corresponding constraint satisfaction problem in the following way: the set of variables $V$ will be the set
\[V=\{(a,b,c) \, |\, a,b,c \in A\}\]
of all triples of elements from $A$. The domain $D$ for every variable will be $A$, the domain of the relational structure $\mathbf{A}$. In order to define the set of constraints $\mathcal{C}$, to every $l$-ary relation $R\in \mathcal{R}$ ($l\leq k$), we associate an $l$-ary relation $\bar{R}\subseteq V^l$, in the following way:
\[((a_1,b_1,c_1),(a_2,b_2,c_2),\ldots,(a_l,b_l,c_l))\in \bar{R}\]  if, and only if \[(a_1,a_2,\ldots,a_l),(b_1,b_2,\ldots,b_l),(c_1,c_2,\ldots,c_l)\in R.\]
To  every $l$-tuple of triples $\bar{x}=((a_1,b_1,c_1),(a_2,b_2,c_2),\ldots,(a_l,b_l,c_l))$ that appears in one of the relations $\bar{R}$, with $\bar{R}$ being $l$-ary $(l\leq k)$,  we associate an $l$-ary constraint $C_{\bar{x}}$, whose scope is the tuple $\bar{x}=((a_1,b_1,c_1),(a_2,b_2,c_2),\ldots,(a_l,b_l,c_l))$ and with the associated $l$-ary constraint relation $R$.  The set of all constraints constructed in this way will be $\mathcal{C}$.

\begin{clm}\label{polyminst} The set of all solutions of the instance $\mathcal{P}=(V,A,\mathcal{C})$, constructed as above, will be the set of all ternary polymorphisms of the relational structure $\mathbf{A}$.
\end{clm} 

Ultimately,  our goal is to construct an algorithm which, when given a finite relational structure $\mathbf{A}$, with $|A|=n$, decides in time $\mathcal{O}(p(n))$, where $p$ is a fixed polynomial of bounded degree, whether $\mathbf{A}$ admits a Maltsev polymorphism and, ideally, gives an explicit definition of such a polymorphism.

First, we make some observations regarding the sizes of the components of $\mathcal{P}$ when $|A|=n$.  Clearly $|V|=|A^3|=n^3$. Also, the number of constraints in $\mathcal{C}$ is bounded above by the number of $l$-tuples in $V$ which can appear in the relations $\bar{R}$ ($R\in\mathcal{R}$).  Each relation $R\in \mathcal{R}$ can contain at most $|A|^l=n^l$ tuples and the total number of tuples in a relation $\bar{R}$ is bounded above by $(n^l)^3$. Thus,  if $|A|=n$, the size of the instance $\mathcal{P}$, whose solutions correspond to the set of all ternary polymorphisms of $\mathbf{A}$ is still polynomial in $n$.

The next step is to introduce additional constraints, whose purpose, is to restrict the set of solutions of $\mathcal{P}$ to precisely those polymorphisms which are Maltsev, i.e. which satisfy
\[m(a,a,b)=m(b,a,a)=b, \quad \mbox{ for all } a,b\in A.\]

To this end, for every variable $i\in V$, where $i$ is either $(a,a,b)$ or $(b,a,a)$, for some $a,b\in A$, we introduce a unary constraint $P_i$, such that $P_i=\{b\}$, and add all such unary constraints $P_i$ $(i=(a,a,b) \mbox{ or } (b,a,a), \, a,b\in A)$ to $\mathcal{C}$. We will abuse the notation and keep referring to this expanded set of constraints as $\mathcal{C}$, and to the resulting instance as $\mathcal{P}$, even though additional unary constraints have been added.

The solutions to this altered instance $\mathcal{P}$ are precisely the Maltsev polymorphisms of $\mathbf{A}$. To see this, if $f$ is a solution of the new instance $\mathcal{P}$, it is a polymorphism of $\mathcal{A}$ by Claim \ref{polyminst}. In addition,  $f$ must satisfy all new unary constraints $P_i$, for all $i$ of the form $(a,a,b)$ or $(b,a,a)$, where $a,b\in A$,  which ensures that it is Maltsev.  Conversely, it is easily seen, that every Maltsev polymorphism $m:A^3\rightarrow A$ is a solution of $\mathcal{P}$.

Before concluding this subsection, we state the following claim, whose proof follows directly from the definitions of products of algebraic and relational structures, as well as the definition of a polymorphism of a relational structure:

\begin{clm} Let $\mathbf{A}$ be a finite relational structure. Then, for any $k$ ($k\geq 1$), $\mathbf{A}^k$ has a Maltsev polymorphism if, and only if $\mathbf{A}$ has a Maltsev polymorphism.
\end{clm}

Our next goal is to use this observation, in order to find a suitable $k$, $(k\geq 1)$, such that the solvability of $\mathcal{P}$ will be reduced to the solvability of an equivalent constraint satisfaction of problem, whose domain will be $A^k$, while all constraints are binary.

\subsection{Reduction of the general CSPs to Binary Relational Structures}\label{sec:reduction}

In this subsection, we outline a reduction of an instance $\mathcal{P}$ of $\operatorname{CSP}(\mathbf{A})$, where $\mathbb{A}$ is a finite idempotent algebra, to a binary instance over a binary relational template, parametrized by $\mathbb{A}^m$, for some $m\geq 1$. The construction is due to L. Barto and M. Kozik and we largely adhere to their exposition in \cite{b-k1}. The reduction is  given by first-order (in fact, quantifier-free) formulas in a bounded number of variables, and can be carried out in deterministic logarithmic space.

An instance $\mathcal{P}$ is said to be \emph{syntactically simple} if it satisfies the following conditions:

\begin{itemize}
\item Every constraint in $\mathcal{C}$  is binary and its scope is a pair of  variables $(x,y)\in V^2$.
\item For every pair of variables $x,y$, there is precisely one constraint $E_{x,y}$ with the scope $(x,y)$.
\item If $x=y$, then $E_{x,x}=\{(a,a)\,\vert\, a\in P_x\}$, where $P_x$ is the $x$-th domain.
\item If $(x,y)$ is the scope of $E_{x,y}$, then $(y,x)$ is the scope of the constraint $E_{y,x}=\{(b,a) \, \vert\, (a,b)\in E_{x,y}\}$ (\emph{symmetry of constraints}).
\end{itemize}

Given any finite algebra $\mathbb{A}$ parameterizing the instance $\mathcal{P}$ such that the maximal arity of a relation in $\mathcal{P}$ is $p$, we define a new, syntactically simple instance $\mathcal{P}$ in the following way:

\begin{itemize} 
\item The instance is parametrized by $\mathbb{A}^{\lceil \frac{p}{2}\rceil}$, which is an algebra satisfying all term identities (equations) $s\approx t$, satisfied by $\mathbb{A}$. 
\item For every $\lceil\frac{p}{2}\rceil$-tuple of variables in $\mathcal{P}$, we introduce a new variable in $\mathcal{P}$ and, if $x=(x_1,\ldots,x_{\lceil\frac{p}{2}\rceil})$ and $y=(y_1,\ldots,y_{\lceil\frac{p}{2}\rceil})$ with $x\neq y$, we introduce a constraint 
\begin{align*}
E_{x,y} &=\{((a_1,\ldots,a_{\lceil\frac{p}{2}\rceil}),(b_1,\ldots,b_{\lceil\frac{p}{2}\rceil}))\,\vert \, (a_1,\ldots,a_{\lceil\frac{p}{2}\rceil},b_1,\ldots,b_{\lceil\frac{p}{2}\rceil})\\
&\mbox{ is a $p$-assignment of values which satisfies all  atomic formulas }\\
& \mbox{ on the tuples of variables } x,y \}
\end{align*}
while, if $x=y$, the relation $E_{x,x}$ is simply the equality of $\lceil\frac{p}{2}\rceil$-tuples in $\mathbb{A}^{\lceil\frac{p}{2}\rceil}$.
\end{itemize}

The binary instance $\mathcal{P}$ constructed in this way will have a solution if, and only if, the instance $\mathcal{P}$ has a solution. 

From the reduction described above, it is easily seen that, if $\mathcal{P}$ is an instance parametrized by a finite algebra $\mathbb{A}$, then the constructed, syntactically simple binary instance $\mathcal{P}$ can be parametrized by the direct product $\mathbb{A}^{\lceil\frac{p}{2}\rceil}$. In particular, if the original instance $\mathcal{P}$ is parametrized by a Maltsev algebra, then so is $\mathcal{P}$.

Therefore, we may assume that, for the given relational structure $\mathbf{A}= (A; \mathcal{R})$ all relations in $\mathcal{R}$ are binary ones. In light of that remark, all constraints in the set of constraints $\mathcal{C}$ of the instance $\mathcal{P}$, constructed in Subsection \ref{sec:polym} can also be assumed to be binary. This assumption will be a key ingredient in the construction of an algorithm for solving the instance $\mathcal{P}$.

\subsection{Symmetric Linear Datalog and (1,2)-Consistency}

A \emph{Datalog program} for a relational template $\mathbf{A}$ is a finite set of rules of the form $T_0\leftarrow T_1,T_2,\ldots, T_n$ where $T_i$'s are atomic formulas. 
$T_0$  is the \emph{head} of the rule, while $T_1,T_2,\ldots, T_n$ form the \emph{body} of the rule.
Each Datalog program consists of two kinds of relational predicates:
the \emph{intentional} ones (IDBs), which are those occurring at least once in the head of some rule, and which are not part of the original signature of the template (they are derived by the computation).
The remaining predicates are said to be the \emph{extensional} ones, or EDBs. They are relations from the signature of the template and do not change during computation; i.e. they cannot appear in the head of any rule.
In addition to those, there is one special, designated IDB, which is nullary (Boolean) and referred to as the \emph{goal} of the program. The semantics of Datalog programs are generally defined
in terms of fixed-point operators. 

 A rule $T_0\leftarrow T_1,T_2,\ldots, T_n$ is said to be \emph{linear} if at most one atomic formula in its body is an IDB. A Datalog program is said to be \emph{linear} if all its rules are linear. The evaluation of a linear Datalog program is in nondeterministic logspace since, from the computational complexity point of view, it reduces to repeated connectivity checks in a finite directed graph corresponding to the program. The linear rules in which an IDB appears in the body are said to be \emph{recursive}.

The \emph{symmetric complement} of a recursive linear rule  $T_0\leftarrow T_1,T_2,\ldots, T_n$ in which, without loss of generality,  the IDBs are $T_0$ and $T_1$, is defined to be the rule
\[T_1\leftarrow T_0,T_2,\ldots, T_n.\] If the rule is non-recursive, its symmetric complement is the rule itself. A linear Datalog program is \emph{symmetric}, if the symmetric complement of every rule also appears in the program.

Given a CSP instance $\mathcal{P}=(V, A, \mathcal{C})$ over a relational template $\mathbf{A}$, its \emph{canonical Symmetric}  (1,2)-\emph{Datalog program} has a unary IDB, $P_i(x)$, for each domain $P_i\subseteq A$ in the instance, with  the said IDBs being the only IDBs in the program. The program allows derivation rules with the body involving at most two variables along with their symmetric complements. 1 indicates that the program is deriving facts about unary relations only, while $2$ indicates that the maximal number of distinct variables in any rule is 2.  

If the instance $\mathcal{P}$ is such that:

\begin{enumerate}
\item all constraints are binary; and
\item for every pair of variables $i,j\in V$, there exists a unique binary constraint $E_{i,j}\in \mathcal{C},$ so that $E_{i,j}=E_{j,i}^{-1}$
\end{enumerate}

if the canonical symmetric (1,2)-Datalog program does not derive a contradiction (empty instance), all constraints  in the derived instance will be subdirect products. This is due to the fact that the only recursive rules involving unary IDBs $R_i, R'_i$ and $R_j, R'_j$, defined  on $P_i$ and $P_j$, respectively, are of the form $R_j (x)\leftarrow E_{i,j}(y,x),R_i(y)$, $R'_j(x)\leftarrow E_{j,i}(x,y),R'_i(y)$, along with their symmetric complements.

Finally, the complexity of computing the canonical symmetric (1,2)-Datalog program for any (not necessarily binary) instance $\mathcal{P}$ of $\operatorname{CSP}(\mathbf{A})$ is in the deterministic logspace (by \cite{reingold2008undirected}) since the run of the program can be viewed as a sequence of connectivity checks in an  undirected graph, associated with $\mathcal{P}$.

\section{The Algorithm}

Throughout this section, we assume that $\mathcal{P}=(V, A, \mathcal{C})$ is an instance of a constraint satisfaction problem over a finite relational template $\mathbf{A}$, where all the constraints in $\mathcal{C}$ are binary, such that $\mathcal{P}$ has a solution if, and only if, $\mathbf{A}$ admits a Maltsev polymorphism. Based on the exposition in Subsection \ref{sec:reduction}, we may assume that for any pairs of distinct indices, $i\neq j$, $i,j\in V$, there are binary constraints $E_{i,j}$ and $E_{j,i}$, such that $E_{i,j}= E_{j,i}^{-1}$. For the sake of simplicity, we will assume that $V=\{1,2,\ldots, N\}$, where $N=n^3$, for $|A|=n$.



The construction of the algorithm was largely inspired by the algorithms for solving Maltsev CSPs in \cite{delic2025constraint} and \cite{delic2026maltsev}. The main difference is that in \cite{delic2025constraint} and \cite{delic2026maltsev}, the algebra of polymorphisms of the underlying finite relational structure was determined to the extent it was necessary to formulate the corresponding algorithms, through given Taylor and Maltsev polymorphisms, respectively.

Here, we are faced with the problem that no nontrivial polymorphisms of the relational structure are known at the outset. Therefore, it is impossible to base the algorithm on the knowledge of, for example, 2-generated algebras or their congruences, or the exhaustive list of minimal nontrivial subalgebras.. Therefore, our algorithm will necessarily be based on the properties of the relational structure, without relying on the existence of particular polymorphisms of $\mathbf{A}$. 

However, based on Proposition \ref{rect}, if a solution to the instance $\mathcal{P}$ exists, i.e. if $\mathbf{A}$ admits a Maltsev polymorphism, every binary constraint $E_{i,j}$ ($i,j\in V$, $i\neq j$), must satisfy the rectangularity property. That is, for any pair of distinct indices $i,j\in V$, if $a,b\in P_i$ and $c,d\in P_j$, where $a\neq b$ and $c\neq d$, if $(a,c), (a,d), (b,d)\in E_{i,j}$, then $(b,c)\in E_{i,j}$.

Thus, the first, preparatory step of the algorithm is to verify that this property holds for all distinct indices $i,j\in V$, if $a,b\in P_i$ and $c,d\in P_j$, where $a\neq b$ and $c\neq d$. There are $N$ indices in $V$ and $n(n-1)/2$ such pairs of distinct elements in every domain, thus all checks of rectangularity for the instance $\mathcal{P}$ can be completed in time that is polynomial in $n$.

If, for any two pairs of distinct elements from two distinct domains, the rectangularity check fails, the algorithm terminates with the output "$\mathbf{A}$ admits no Maltsev polymorphism." Therefore, we assume that all rectangularity checks succeed and we proceed with the formulation of the algorithm.

\subsection{The Algorithm in the Rectangular Case}

As mentioned earlier, when trying to formulate an algorithm which would solve the instance $\mathcal{P}$ of the CSP whose template is a finite relational structure $\mathbf{A}$, all of whose relations are binary, the lack of knowledge of the algebraic structure on $A$, induced by the polymorphisms of $\mathbf{A}$, seems to be a significant obstacle that needs to be overcome. However, under the assumption of rectangularity, some of the algebraic structure can be recovered in the form of certain congruences, which can be defined exclusively from the binary relations of $\mathbf{A}$, i.e. from the constraints in $\mathcal{C}$.

Let $\mathbb{S}\leq \mathbb{B}\times \mathbb{C}$ be a subdirect product of two algebras of the same type, such that $\mathbb{S}$ is rectangular. We define a binary relation $\theta_S^{\mathbb{B}}\subseteq B\times B$ in the following way:
\[(b,b')\in \theta_S^{\mathbb{B}} \mbox{ if, and only if } \exists c\in C \mbox{ so that } (b,c),(b'c)\in S.\]

\begin{clm}  $\theta_S^{\mathbb{B}}$ is an equivalence relation on $B$. Furthermore, $ \theta_S^{\mathbb{B}}$ is a congruence of $\mathbb{B}$.
\end{clm}

\begin{proof} The reflexivity and symmetry of $ \theta_S^{\mathbb{B}}$ follow directly from the definition of the binary relation and the fact that $\mathbb{S}$ is a subdirect product.. Since $\mathbb{S}$ is rectangular, it is easily shown that  $\theta_S^{\mathbb{B}}$ is also transitive.

The relation  $\theta_S^{\mathbb{B}}$ is defined by a positive-primitive formula and is, therefore, compatible with all polymorphisms of $\mathbb{B}$.
\end{proof}

The relation $ \theta_S^{\mathbb{C}}$ can be defined on $C$ analogously, and the proof that it is a congruence of $\mathbb{C}$ is  identical.

\begin{cor} If $\mathbb{S}\leq \mathbb{B}\times \mathbb{C}$ is a rectangular subdirect product of two algebras $\mathbb{B}$ and $\mathbb{C}$ of the same type, which are both simple, then $\mathbb{S}$ is either:
\begin{itemize}
\item[(a)] the graph of an isomorphism $\phi:\mathbb{B}\rightarrow\mathbb{C}$; or
\item[(b)] $\mathbb{S}=\mathbb{B}\times\mathbb{C}$.
\end{itemize}
\end{cor}

Let $\mathbb{S}\leq \mathbb{B}\times\mathbb{C}$ be a rectangular subdirect product, and  $\theta_S^{\mathbb{B}}$ and  $\theta_S^{\mathbb{C}}$ be the associated congruences on $\mathbb{B}$ and $\mathbb{C}$, respectively.  Let $\tilde{B}$ and $\tilde{C}$ be the congruence classes of $ \theta_S^{\mathbb{B}}$ and $ \theta_S^{\mathbb{C}}$, respectively, so that, for some $b\in B$ and $c\in C$, $(b,c)\in S$. Then, due to the rectangularity of $\mathbb{S}$
\[S\cap (\tilde{B}\times\tilde{C})=\tilde{B}\times\tilde{C}.\]

Thus, any rectangular subdirect product 
\[\mathbb{S}\leq \mathbb{B}\times\mathbb{C}\]
of two algebras $\mathbb{B}$ and $\mathbb{C}$ of the same type, can be represented as a disjoint union of the sets of the form $\tilde{B}\times\tilde{C}$, where $\tilde{B}$ and $\tilde{C}$ are congruence classes of $ \theta_S^{\mathbb{B}}$ and $ \theta_S^{\mathbb{C}}$, respectively.

\subsection{Triple Graph of a Syntactically Simple Binary Instance}\label{graph}

Let $\mathcal{P}$ be a syntactically simple binary instance of $\operatorname{CSP}(\mathbb{A}')$, where $\mathbb{A}'$ is a finite Maltsev algebra. To every such instance, we can associate an undirected graph $\mathbf{G}_{\mathcal{P}}$, defined in the following way:

\begin{itemize}
\item the vertices of $\mathbf{G}_{\mathcal{P}}$ are the triples $(i,a,b)$, such that $i\in V$ and $a,b\in P_i$, $a\neq b$.
\item let $a,b\in P_i$, ($a\neq b$) and $c,d\in P_j$ ($c\neq d$), with $i\neq j$. Then, $(i,a,b)$ and $(j,c,d)$ will form an edge  in $\mathbf{G}_{\mathcal{P}}$ if, and only if the following is true:
\[(a,c),(b,d)\in E_{i,j}, \quad \mbox{ and } \quad (a,d),(b,c)\not\in E_{i,j}.\]
[or, equivalently, $(c,a),(d,b)\in E_{j,i}$, and  $(d,a),(c,b)\not\in E_{j,i}.$]
\end{itemize}

From the definition of the edge relation in $\mathbf{G}_{\mathcal{P}}$,  it is easily seen that $(i,a,b)$ and $(j,c,d)$ form an edge in $\mathbf{G}_{\mathcal{P}}$ if, and only if, $(i,b,a)$ and $(j,d,c)$ form an edge. For that reason, we can identify the vertices $(i,a,b)$ and $(i,b,a)$, for any $i\in V$ and $a,b\in P_i$, with $a\neq b$. In the remainder of the paper, we will tacitly assume that such an identification has been made.

In addition to the instance $\mathcal{P}$, we will assume that its associated graph $\mathbf{G}_{\mathcal{P}}$ is also provided to the algorithm as a part of the input. The graph $\mathbf{G}_{\mathcal{P}}$ can be constructed in logspace, given the instance $\mathcal{P}$. If $(i,a,b)$ is a vertex of $\mathbf{G}_{\mathcal{P}}$, its connected component in this triple graph will be denoted $C(i,a,b)$.

We point out that every edge of the graph $\mathbf{G}_{\mathcal{P}}$ represents a subdirect product of two 2-generated subuniverses of the Maltsev algebra $\mathbb{A}'$. In particular, every edge in the triple graph represents an unlinked subdirect product of two such algebras. In the case when $\operatorname{Sg}_i(a,b)$ and $\operatorname{Sg}_j(c,d)$ are both simple algebras, the edge between $(i,a,b)$ and $(j,c,d)$ implies the existence of an isomorphism between the two algebras mapping $a$ to $c$ and $b$ to $d$.

The motivation for the definition of the triple graph associated to a syntactically simple binary instance  was provided by \cite{egri2014space}, where a similar definition of the notion of a triple digraph associated with a conservative digraph was used to prove that CSPs over conservative digraph templates which admit Hagemann-Mitschke terms can be solved in deterministic logspace.

\subsection{Group of a Triple Graph Vertex}

Let $(i,a,b)\in L$ be a triple, with $a\neq b$. The \emph{group} $J_{i,a,b}$ of $(i,a,b)$, is defined in the following way: let $j\in V$  be such that, for some $(j,c,d)\in P_j$, $(c\neq d)$, there exists a path in the triple graph from $(i,a,b)$ to $(j,c,d)$. Then, $j\in J_{i,a,b}$. From the fact that the existence of a path in an undirected graph can be verified in deterministic logspace, we get the following proposition.

\begin{prop} For every $i\in V$, $a, b\in P_i$ $(a\neq b)$, the group $J_{i,a,b}$ of $(i,a,b)$ can be computed in deterministic logspace. 
\end{prop}

We also assume that, for every $(i,a,b)$, where $a\neq b$, $i\in J_{i,a,b}$.

The notion of the group of a vertex $(i,a,b)$ of the triple graph associated with the instance $\mathcal{P}$ will play the crucial role in the construction of the algorithm for solving $\mathcal{P}$.  We start with with the following observation:

\begin{clm}\label{minimal} If at least two solutions to $\mathcal{P}$ exist, then, there is a subdirect product $\mathbb{S}'$ of trivial and minimal subalgebras of $\mathbb{A}=(A; \operatorname{Pol}(\mathbf{A}))$ such that every $N$-tuple in $S'$ is a solution.
\end{clm}

\begin{proof} Let $\mathbb{S}$ be the subdirect product consisting of all solutions to $\mathcal{P}$. Then,
\[\mathbb{S}\leq \mathbb{P}'_1\times\mathbb{P}'_2\times\cdots\times \mathbb{P}'_N,\]
where, for $1\leq i\leq N$, $\mathbb{P}'_i\leq \mathbb{P}_i$.  Clearly, every nontrivial $\mathbb{P}'_i$ contains a minimal subuniverse $\tilde{\mathbb{P}}_i$.  If $i_0$ is a fixed index for which $\mathbb{P}'_i$ is nontrivial, we fix the corresponding $\tilde{\mathbb{P}}_{i_0}$, restrict the $i_0$-th domain to it, and apply the (1,2)-symmetric Datalog program to it. The output will be nonempty, since $\mathbb{S}$ is the solution set of $\mathcal{P}$. Proceeding inductively, we can reduce the domains to trivial and minimal subalgebras of $\mathbb{A}$.
\end{proof}

Therefore, we will design our algorithm in such a way that it verifies consistencies in subinstances of $\mathcal{P}$, whose domains are either trivial or minimal subalgebras of $\mathbb{A}$. Again, a potential obstacle is the lack of explicit knowledge of such minimal subalgebras since the algebraic structure is not given to us. However, as will become clear later, rectangularity of binary constraints will be extremely helpful.

So, under the assumption of rectangularity of binary constraints, let us examine the meaning of the group of $(i,a,b)$, for some $a\neq b$.  In particular, what is the significance of $J_{i,a,b}$, when the subuniverse of $\mathbb{A}$, generated by $a$ and $b$, $\operatorname{Sg}_{\mathbb{A}}(a,b)$, is minimal and, therefore, simple? Suppose $j\in V$ is such, that there exists $c\in P_j$, such that $(a,c),(b,c)\in E_{i,j}$. Then,  for every $\mathbb{D}\leq \mathbb{P}_j$ and $\mathbb{S}$ such that 
\[\mathbb{S}\leq\operatorname{Sg}_{\mathbb{A}}(a,b)\times \mathbb{D},\]
$\mathbb{S}$ must be the full product $\operatorname{Sg}_{\mathbb{A}}(a,b)\times \mathbb{D}$. Namely, for such $\mathbb{S}$, we consider the congruences $\theta_S^{\operatorname{Sg}_\mathbb{A}(a,b)}$ and $\theta_S^{\mathbb{D}}.$ Since the former is the full relation on $\operatorname{Sg}_\mathbb{A}(a,b)$, the same must be true of $\theta_S^{\mathbb{D}}$ on $\mathbb{D}$.

Now, if there exists a path $\alpha$ in the triple graph from $(i,a,b)$ to $(j,c,d)$, then if $\mathbb{D}$ is a minimal subuniverse of $\mathbb{P}_j$ generated by $c$ and $d$, then the path $\alpha$ will induce an isomorphism from $\operatorname{Sg}_{\mathbb{A}}(a,b)$ to $\mathbb{D}$.  For that reason, we can view the edges of the triple graph as isomorphisms in the case when the vertices $(j,c,d)$, $c\neq d$ which appear in the path where every $j\in J_{i,a,b}$ is in the group of the fixed vertex $(i,a,b)$ and all $c$ and $d$ are such that $\operatorname{Sg}_{\mathbb{A}}(c,d)$ are minimal subalgebras of $\mathbb{A}$.

Let us now assume that we are given a (1,2)-consistent subinstance $\mathcal{P}'$ of $\mathcal{P}$, all of whose domains are either trivial or minimal subalgebras of $\mathbb{A}$. Since $\mathcal{P}'$ is (1,2)-consistent,  the restriction of every binary constraint $E_{i,j}\in \mathcal{C}$ ($i\neq j$) to $\mathcal{P}'$, which we will denote $E'_{i,j}\in \mathcal{C}$ is a subdirect product. As mentioned earlier, every such subdirect product is either a graph of an isomorphism between two minimal subuniverses, or their full direct product. In trying to determine the necessary consistency requirements, we can disregard the binary constraints which are full direct products, since they do not affect the existence of solutions to $\mathcal{P}'$.

The set of variables $V=\{1,\ldots,N\}$ can be partitioned into $J_{i_l,a_l,b_l}$, for some $l=1,\ldots,m$.  Based on an earlier remark, between any two such groups, all binary constraints are full products. Therefore, to ensure that $\mathcal{P}'$ has a solution, it is sufficient to show that every subinstance of $\mathcal{P}'$, in which the set of variables $V$ is restricted to a subset consisting of the group $J_{i_l,a_l,b_l}$ has a solution.  We will now try to analyze what path consistency conditions such a subinstance must satisfy in order for a solution to exist.  

Let us fix one such group, and denote it $J_{i_0,a_0,b_0}$. If a solution exists through $a_0$, say, then, for every closed path of variable indices from $J_{i_0,a_0,b_0}$, which starts and ends in $i_0$, there must be a path in the multiconsistency graph of the instance $\mathcal{P}'$, which starts and ends in $a_0$.  Otherwise, since the instance $\mathcal{P}'$ is (1,2)-consistent, there is a path from $a_0$ to some vertex $c\in \operatorname{Sg}_{\mathbb{A}}(a_0,b_0)$, where $c\neq a$. Since a minimal algebra is generated by any pair of distinct elements, the solution can only exist when the path in the multiconsistency graph corresponding to the simple path of variables starting and ending in $i_0$, which starts at $a_0$, cannot contain any other element of $\operatorname{Sg}_{\mathbb{A}}(a_0,c)$, for any $c\in\operatorname{Sg}_{\mathbb{A}}(a_0,b_0)$, $c\neq b_0$.

Thus, in order to ensure consistency (and the existence of a solution) through $a_0$ in $\mathcal{P}'$, the subinstance of $\mathcal{P}'$, obtained by restricting the $i_0$-th domain to $\{a_0\}$, must pass the (1,2)-symmetric Datalog test.  We will base the construction of the algorithm on this type of consistency check. The idea behind the construction is to test all possible minimal subuniverses of the domains $\mathbb{P}_1,\mathbb{P}_2,\ldots,\mathbb{P}_N$ of $\mathcal{P}$,  which are generated by two distinct elements $a,b\in A$,  using the groups $J_{i,a,b}$ of $V$ ($i=1,\ldots,N$) of the triple graph, in the manner described above.

\subsubsection{$(i,a,b)$-Test}

At the start, we form a list $L=\{(i,a,b)\, : \, i\in[N], a,b \in P_i\}$, where we allow the possibility that $a=b$.

If $a=b$, the $(i,a,b)=(i,a,a)$, where $a\in P_i$, reduces the $i$-th domain of $\mathcal{P}$ to $\mathbb{P}_i'=\{a\}$, while $\mathbb{P}_j'=\mathbb{P}_j$, for $j\neq i$, and runs the (1,2)-symmetric Datalog program on this new subinstance $\mathcal{P}'$. If the test results in an empty instance, there cannot be any solution through $a$ in $P_i$ and all triples of the form $(i,a,b)$, for some $b\in P_i$ are removed from $L$. Also, $a$ is removed from $P_i$.

Let $(i,a,b)\in L$ be a triple, with $a\neq b$. The $(i,a,b)$-test starts by constructing the group $J_{i,a,b}$ of $(i,a,b)$. As mentioned earlier, this step can be completed within deterministic logspace, and is, therefore, in polynomial time.

\begin{prop} For every $i\in V$, $a, b\in P_i$ $(a\neq b)$, the group $J_{i,a,b}$ of $(i,a,b)$ can be computed in deterministic logspace. 
\end{prop}

After the group $J_{i,a,b}$ of $(i,a,b)$ has been computed, along with the restriction of the triple graph to the domains in $J_{i,a,b}$,  the next steps consist of the following:

\begin{enumerate}
\item   $P'_i$ is $P_i$ restricted to $\{a\}$;
\item if $j\in J_{i,a,b}$ but $j\neq i$, let $\alpha$ be a path from $(i,a,b)$ to $(j,c,d)$, for some $c,d\in P_j$, such that $c\neq d$. That is, we fix a path $\alpha_V$, which is the projection to the first coordinate (variables in $V$)  of a path $\alpha$ in the triple digraph, which witnesses the fact that $j\in J_{i,a,b}$. We define $P_j'$ to consist of \emph{all} $c\in P_j$ such that there is an $\alpha_V$-path in the first coordinate  in the triple graph from $(i,a,b)$ to $(j,c,d)$, for some $d\in P_j$, $c\neq d$;
\item if $j\not\in J_{i,a,b}$, we let $P_j'=P_j$;
\item Run the (1,2)-symmetric Datalog program on the instance $\mathcal{P}'$, in which, $P'_i$ are defined as in (1)-(3). If the program results in an empty instance, $(i,a,b)$-test fails and $(i,a,b)$ is removed from $L$; otherwise
\item Perform steps (1)-(4) with the roles of $a$ and $b$ reversed.
\end{enumerate}

We iterate $(i,a,b)$-tests for the triples in $L$ until the list $L$ does not change anymore. Clearly, this process can proceed for at most as many steps as there are triples in $L$, and that list contains polynomially many (in $n=|A|$) such triples. If, at the end of this process, for some $i\in V$, there are no triples of the form $(i,a,b)$, the $i$-th domain of the reduced instance is empty, and the instance $\mathcal{P}$ has no solution. Otherwise, we will show that a solution exists.


\subsection{Proof of Correctness}

To complete our proof, we show that if, at the end of iterated $(i,a,b)$-tests, for every $i\in V$, there is some $(i,a,b)\in L$, is equivalent to the CSP having a solution. 

If, for every $i\in V$, there is a single triple $(i,a_i,a_i)\in L$, for some $a_i\in A$, then, since every $(i,a_i,a_i)$-test succeeds, we have a (1,2)-consistent instance with all domains being singletons. In that case, it is immediately clear that $f:V\rightarrow A$, such that $f(i)=a_i$ is a solution to $\mathcal{P}$.

If, for every $i\in V$, there is a single triple in L, of the form $(i,a,b)$, where either a=b, or $\operatorname{Sg}(a,b)$ is a minimal subalgebra of $\mathbb{A}$, then we have a (1,2)-consistent subinstance $\mathcal{P}'$ of $\mathcal{P}$, all of whose domains are either trivial or minimal subalgebras of $\mathbb{A}$. The set of variables $V$ of $\mathcal{P}'$ is partitioned into groups $J_{i_0,a_0,b_0},J_{i_1,a_1,b_1},\ldots,J_{i_k,a_k,b_k}$ of some triples $(i_0,a_0,b_0),(i_1,a_1,b_1),\ldots,(i_k,a_k,b_k)\in L$. All the domains corresponding to the variables in the same group are isomorphic; either they are all trivial, or isomorphic to a particular minimal subalgebra of $\mathbb{A}$. The success of the $(i_l,a_l,b_l)$-tests, for $0\leq l\leq k$ implies the existence of a partial solution over every group $J_{i_l,a_l,b_l}$, for $0\leq l\leq k$.  Since the restrictions of binary constraints to domains whose indices are from different groups are direct products, a solution exists over the entire set of variables $V$.

We define a \emph{path pattern} in $V$ to be a sequence $\pi=i_0,i_1,i_2,\ldots,i_k$ of variables from $V$. If all pairs of variables in $\pi$, except, possibly $i_0$ and $i_k$ are distinct, we say that the path pattern $\pi$ is \emph{simple}. If $i_0=i_k$, the path pattern is said to be \emph{closed}.

Given a path pattern $\pi$ in $V$, we say that a \emph{realization} of $\pi=i_0,i_1,i_2,\ldots,i_k$ in $\mathcal{P}'$ is a mapping $f:V\rightarrow A$, such that $f(i_l)=a_l$, for $l=0,1,\ldots,k$, where $a_l\in P'_l$ and $(a_l,a_{l+1})\in E_{l,l+1}$, for $l=0,1,\ldots,k-1$. If the path pattern $\pi$ is closed, then we say that a realization $f$ is \emph{strong} if $a_0=a_k$.

Now, let $i\in V$ and $(i,a,b)\in L$. Then,

\begin{clm} For every $c,d\in A$ such that $\operatorname{Sg}_\mathbb{A}(c,d)\subseteq \operatorname{Sg}_\mathbb{A}(a,b)$, if $(i,a,b)\in L$ then $(i,c,d)\in L$.
\end{clm}

\begin{proof}This claim follows directly from the following observation: $J_{i,c,d}\subseteq J_{i,a,b}$. 

To see this, notice that if $j\not\in J_{i,a,b}$, for every path pattern $\pi=i,j_1,\ldots,j_k=j$, there exist two realizations of $\pi$, one starting at $a$, the other one at $b$, and both ending in $e$, for some $e\in P_j$. Since all polymorphisms of $\mathbf{A}$ (i.e. the operations of $\mathbb{A}$) are idempotent, there will exist realizations of $\pi$ with values at $i$ being $c$ and $d$, respectively, and the value at $j$ being $e$, so $j\not\in J_{i,c,d}$.

A (1,2)-symmetric Datalog program is defined using positive primitive formulas. So, if the $i$-th domain $P'_i$ is restricted to any element $c\in \operatorname{Sg}_\mathbb{A}(a,b)$, from the fact that such a program suceeds when the $i$-th domain is restricted to $\{a\}$ and $\{b\}$, the fact that $J_{i,c,d}\subseteq J_{i,a,b}$ and the positive primitive definition of the program clauses, the program will succeed for the restriction to any $c\in \operatorname{Sg}_\mathbb{A}(a,b)$.

Thus, $(i,c,d)$-test succeeds and $(i,c,d)\in L$.
\end{proof}

We will show that the following consistency condition, which ensures the solvability of $\mathcal{P}'$ in the case when all the domains of the (1,2)-consistent instance $\mathcal{P}'$ are either trivial or minimal subalgebras of $\mathbb{A}$ , is preserved in all (1,2)-consistent subinstances of $\mathcal{P}'$:
\bigskip

\begin{itemize}
    \item[(*)]for every $i\in V$, $a,b\in P_i'$, where $a\neq b$ and $\operatorname{Sg}_{\mathbb{A}}(a,b)$ is a minimal subalgebra of $\mathbb{A}$, and every closed simple path pattern $\pi$, whose variables are in $J_{i,a,b}$, and every $c\in\operatorname{Sg}_{P'_i}(a,b)$, there exists a strong realization of $\pi$, whose endpoints are $c$. 
\end{itemize}
 
If $(i,a,b)$-test succeeds, where $\operatorname{Sg}_{P'_i}(a,b)$ is a minimal algebra and $\pi$  a simple closed path pattern in $J_{i,c,d}$, it is also a simple closed path pattern in $J_{i,a,b}$ since $J_{i,c,d}\subseteq J_{i,a,b}$. Now, let $f_a$ and $f_b$ be realizations of $\pi$ whose endpoints are $a$ and $b$, respectively. Since $c$ is in the subuniverse generated by $a$ and $b$, there exists a polymorphism $p(x,y)$ of $\mathbf{A}$, such that $c=p(a,b)$. Then, for every variable $j$ appearing in $\pi$, we define $f_c(j)=p(f_a(j),f_b(j))$. Then, $f_c$ will be a realization of $\pi$, whose endpoints are $c$. This implies that, if the $(i,a,b)$-test succeeds, then the consistency condition holds for all $c\in \operatorname{Sg}_{\mathbb{P'}_i}(a,b)$.

Next, we observe that, if $\mathcal{P}''$ is a subinstance of $\mathcal{P}'$, then, for any $(i,a,b)$, $a\neq b$, such that $a,b\in P''_i$, the group $J_{i,a,b}$ in $\mathcal{P}''$ coincides with the group $J_{i,a,b}$ in $\mathcal{P}'$. This follows directly from the definition of the group of a vertex in the triple graph.

For that reason, if the simple closed path realization condition is satisfied for $a,b\in \mathbb{P}'_i$ in $\mathcal{P}'$, it is also satisfied for both $a$ and $b$ in any proper (1,2)-consistent subinstance $\mathcal{P}''$ of $\mathcal{P}'$. 

Finally, if for some $i\in V$, there exists a triple $(i,a,b)\in L$, with $a\neq b$, such that the subuniverse $\operatorname{Sg}_{\mathbb{A}}(a,b)$  is not minimal, there exists some $(i,c,d)\in L$, where $\operatorname{Sg}_{\mathbb{A}}(c,d)$  is a minimal subalgebra of $\operatorname{Sg}_{\mathbb{A}}(a,b)$. Now, we can consider a smaller subinstance $\mathcal{P}''$ is obtained by restricting $P_i'$ to $\operatorname{Sg}_{\mathbb{A}}(c,d)$. Using an analogous argument as the one used above, one can show that, since $a$ and $b$ are endpoints of strong realizations of all simple closed path patterns in $J_{i,a,b}$, so are $e$ and $f$ for any simple closed path pattern in $J_{i,c,d}$, for any pair of distinct elements $e,f\in\operatorname{Sg}_{\mathbb{A}}(c,d)$. 


\begin{thm}
    If applying the $(i,a,b)$-tests to the instance $\mathcal{P}$ results in a nonempty subinstance $\mathcal{P}'$, $\mathcal{P}$ has a solution and, consequently, the finite relational structure $\mathbf{A}$ has a Maltsev polymorphism.
\end{thm}

\section{Conclusion and Open Problems}

In this article, we developed an algorithm for verifying whether a given finite relational structure admits a Maltsev polymorphism, which operates in polynomial time in the size of the relational structure. 

The algorithm was inspired by a similar algorithm in \cite{delic2026maltsev}, which places the problem of solving an instance of a constraint satisfaction problem over a finite relational template with a Maltsev polymorphism, in the parallel complexity class $\textsc{NC}^2$. More specifically, in (cite the paper again), it was shown that such a problem is in the complexity classes \#\textsc{L} of the problems that can be solved in deterministic logspace with counting. 

By reducing the problem of checking whether a given finite relational structure admits a Maltsev polymorphism to solving an instance of a constraint satisfaction problem over a finite relational structure all of whose relations are binary, we were able to exploit key ideas from (cite again) in the following sense: once all binary constraints are checked for high degree of ``symmetry", necessitated by the existence of a Maltsev polymorphism, the problem becomes very similar to solving systems of equations over elementary Abelian $p$-groups, for certain values of $p$ closely linked to the size of the relational template, where each equation involves two variables.

Such systems of linear equations can be solved using searches in an undirected graph (the triple graph) of the instance in addition to counting. In our case, the main obstacle was the lack of knowledge of \emph{any} explicit polymorphisms of the relational template, which precluded the possibility of analyzing the associated systems of linear equations, since determining minimal algebras of the algebraic structure on the template was not feasible. 

However, using $(1,2)$-symmetric Datalog programs, we were able to verify local (1,2)-consistency to the extent which was necessary to see if a pair of distinct elements could be contained in a minimal subalgebra, which is one of the elementary Abelian $p$-groups that contains solutions to the instance through every one of its elements. 

We believe that further analysis  of this algorithm may be refined to  answer the following question:

\begin{prob} If $\mathbf{A}$ is a finite relational template, is there an algorithm, which verifies whether $\mathbf{A}$ admits a Maltsev polymorphism whose complexity is in a class conjectured to be properly contained in \textsc{P}? In particular, is there such an algorithm in $\textsc{NC}^2$ or, even more specifically, in \#\textsc{L}?
\end{prob}

\bibliography{digraph_reduction} 
\bibliographystyle{alphaurl}

\end{document}